\documentclass[journal]{IEEEtran}
\IEEEoverridecommandlockouts
\usepackage{etex}
\usepackage{amsmath,amssymb}
\usepackage{float}
\usepackage{lipsum}
\usepackage{array}
\usepackage{cite}
\usepackage{textcomp}
\usepackage{hhline}
\usepackage{siunitx}
\usepackage{balance}
\usepackage{makecell}
\usepackage{mathtools}
\usepackage{float}
\usepackage[pdftex]{graphicx}
\usepackage{siunitx}
\usepackage[dvipsnames]{xcolor}
\usepackage{xcolor}
\usepackage{colortbl}
\usepackage{tikz}
\usetikzlibrary{decorations}
\usetikzlibrary{calc} 
\tikzset{>=latex}
\usetikzlibrary{arrows,shapes}
\usetikzlibrary{arrows}
\usetikzlibrary{plotmarks}
\usepackage{ctable}
\usepackage{pgfplots}
\usepackage[english]{babel}
\usepackage{color}
\usepackage[utf8]{inputenc}
\usepackage[T1]{fontenc}
\usetikzlibrary{decorations.pathreplacing,shapes.misc}
\usetikzlibrary{patterns}
\usepackage{tcolorbox}
\usepackage{balance}
\usepackage{subfigure}
\usepackage{hyphenat}
\usepackage{amsmath,bm}

\addto\captionsenglish{}
\newcommand\scalemath[2]{\scalebox{#1}{\mbox{\ensuremath{\displaystyle #2}}}}

\newcommand{\AS}{\textcolor{black}}

\newcommand{\rmT}{\mathsf{T}}
\newcommand{\Bmat}{\boldsymbol{B}}
\newcommand{\opt}{\tilde{l}}
\newcommand{\row}{\mathsf{r}}
\newcommand{\mep}{x}

\renewcommand{\a}{\boldsymbol{a}}
\renewcommand{\b}{\boldsymbol{b}}

\newcommand{\nc}{n}

\newcommand{\kc}{k}

\newcommand{\la}{\boldsymbol{a}}
\newcommand{\lb}{\boldsymbol{b}}

\newcommand{\cc}{\boldsymbol{C}}

\newcommand{\dgmd}{{d}_\mathsf{GD}}

\newcommand{\dGMD}{\dgmd(\la,\lb)}

\newcommand{\BB}{\mathsf{B}}

\newcommand{\mut}{\tilde{\mu}}

\def\forcemath#1{\ifmmode #1 \else $#1$\fi}

\newcommand{\agia}{\textcolor{blue}}

\ifCLASSINFOpdf
 \else
  \fi

\begin{document}


\title{Novel High-Throughput Decoding Algorithms for Product and Staircase Codes based on Error-and-Erasure Decoding}

\author{Alireza Sheikh, \IEEEmembership{Member, IEEE}, Alexandre Graell i
Amat, \IEEEmembership{Senior Member, IEEE}, and \\ Alex Alvarado, \IEEEmembership{Senior~Member,~IEEE} \\ \IEEEauthorblockA{
	\thanks{A. Sheikh and A. Alvarado are with the Department of Electrical Engineering, Eindhoven University of Technology, PO Box 513
	5600 MB Eindhoven, The Netherlands (emails: \{asheikh, a.alvarado\}@tue.nl). The work of A. Sheikh and A. Alvarado has received funding from the European Research Council (ERC) under the European Union's Horizon 2020 research and innovation programme (grant agreement No 757791).}	
    \thanks{A. Graell i Amat is with the Department of Electrical Engineering, Chalmers University of Technology, SE-41296 Gothenburg, Sweden (email: alexandre.graell@chalmers.se).}
}}


\maketitle


\begin{abstract}
	
Product codes (PCs) and staircase codes (SCCs) are conventionally decoded based on bounded distance decoding (BDD) of the component codes and iterating between row and column decoders. The performance of iterative BDD (iBDD) can be improved using soft-aided (hybrid) algorithms. Among these, iBDD with combined reliability (iBDD-CR) has been recently proposed for PCs, yielding sizeable performance gains at the expense of a minor increase in complexity compared to iBDD.
In this paper, we first  extend iBDD-CR to SCCs. We then propose two novel decoding algorithms for PCs and SCCs which improve upon iBDD-CR. The new algorithms use an extra decoding attempt based on error and erasure decoding of the component codes. The proposed algorithms require only the exchange of hard messages between component decoders, making them an attractive solution for ultra high-throughput fiber-optic systems. Simulation results show that our algorithms based on two decoding attempts achieve gains of up to $0.88$~dB for both PCs and SCCs. This corresponds to a $33\%$ optical reach enhancement over iBDD with bit-interleaved coded modulation using $256$ quadrature amplitude modulation.

\end{abstract}



\IEEEpeerreviewmaketitle


\section{Introduction}\label{Sec:Intro}

\IEEEPARstart{T}{he} transmission rate of a single optical fiber has increased by a factor of $8000$ over the past $30$ years \cite{Huawei2016}. Recently, a chipset with the capacity of $800$ Gbit/s/$\lambda$ has become commercially available \cite{ciene2020}. To cope with the growth of applications such as cloud computing, internet-of-things, video-on-demand, etc., next-generation optical line cards target data rates of $1$ Tb/s/$\lambda$ and beyond. Reliable transmission at such high data rates can not be achieved using off-the-shelf digital signal processing (DSP) nor by exclusively relying on the improvement of integrated circuits (due to the end of Moore's law\cite{Steegen17}). Forward error correction (FEC) is an essential component of the receiver DSP and must be adapted accordingly to cope with the current trends on data rate. Designing high-performance FEC decoders for ultra high-speeds is very challenging due to the strict latency and power constraints. 

Product codes (PCs) \cite{Elias1954} and staircase codes (SCCs) \cite{staircase_frank} are popular FECs for high-throughput applications such as fiber-optic systems and have been included in several recommendations (e.g., ITU-T G.709.2/Y.1331.2 \cite{ITUTn} and $400$ZR \cite{400ZR2018}). Soft-decision decoding (SDD) of PCs---also known as turbo product decoding (TPD)---was proposed already more than 20 years ago in \cite{Pyn98}. TPD provides excellent performance at the cost of a high internal decoder data flow due to the iterative exchange of soft messages, which significantly limits the achievable throughput \cite{Schlafer2013}. Further, such soft message passing entails high power consumption, e.g., $10$ W  for $128$ Gbps line rate\cite{Viasat66200}\footnote{The employed code rate is not specified, hence, the corresponding information rate is unknown to the authors.}. An alternative to TPD is to employ iterative hard-decision decoding (HDD). HDD requires much lower decoding complexity (e.g., $0.62$~W for $317$ Gbps information rate \cite{chris2019}), however, it entails large performance losses ($1$--$2$~dB depending on the code rate) compared to SDD. \agia{}

Recently, \AS{several \emph{hybrid} soft-decision decoding (SDD) and hard-decision decoding (HDD)}  architectures have been proposed for both PCs and SCCs, which provide a suitable performance-complexity tradeoff between SDD and HDD \cite{She18b,She19,sheikhTCOM19,Yia2019,GabrieleSABMSR2019,optimal_decsheikh}. The unifying  idea of these schemes is to employ HDD as the decoding core, while exploiting some level of soft information to improve the overall decoder performance.\footnote{Other proposals for high-throughput coding schemes comprise the concatenation of an inner code with SDD and an outer code with HDD\cite{Ksch18}, LDPC codes with binary, ternary, and  quaternary message passing \cite{Richardson2001, Lechner2012, Steiner19}, anchor decoding (AD) of PCs and SCCs based on tracking the conflicts between component decoders \cite{Hag18}, and two-stage decoding based on multi-stage decoding and exploiting both SDD and HDD \cite{Montorsi2018,Montorsi2019}.} In \cite{She18b} and \cite{She19}, two decoding algorithms for PCs were proposed based on error-and-erasure decoding (EDD) \cite{ForneyGMD}. 
In \cite{sheikhTCOM19}, a low-complexity decoding algorithm called iBDD with scaled reliability (iBDD-SR) was presented for both PC and SCCs. SR models the reliability of the hard decisions of the BDD of the component codes by combining the hard decisions properly scaled by a scaling factor and the channel LLRs. In \cite{Yia2019}, the soft-aided bit-marking (SABM) decoder was proposed for PCs and SCCs based on flipping the least reliable bits and a heuristic miscorrection detection procedure. The performance of SABM was further improved for PCs based by combining SABM with the SR principle \cite{GabrieleSABMSR2019}. 

The latest algorithm in the class of \emph{hybrid} decoding architectures was introduced in \cite{optimal_decsheikh}, where iBDD with combined reliability (iBDD-CR) was presented. iBDD-CR improves iBDD-SR by performing the optimal combining (optimal for large block lengths and extrinsic iBDD) of the hard decisions and channel LLRs. This combining is found based on density evolution (DE) analysis of the  generalized low-density parity-check (GLDPC) code ensemble encompassing PCs. Employing the iBDD-CR decoder for SCCs demands finding an accurate estimate of the reliability of the HDD outputs for SCCs, which is not necessarily the same as for PCs.  

This paper extends \cite{optimal_decsheikh} in two different directions. First,  we extend iBDD-CR to SCCs. In particular, we perform a DE analysis for iBDD-CR based on the spatially-coupled GLDPC (SC-GLDPC) code ensemble that contains  SCCs as particular instances. The derived DE provides an accurate estimate of the reliability of the hard decisions of the BDD of the SCC component codes, which is exploited in iBDD-CR. Second, we propose two novel decoding algorithms for PCs and SCCs that enhance the iBDD-CR architecture with EDD of the component codes. The proposed decoders can be efficiently implemented by only exchanging hard messages between component decoders. We perform an algorithmic-level complexity comparison between our proposed schemes and the decoders of \cite{She18b,She19,sheikhTCOM19,Yia2019,GabrieleSABMSR2019,optimal_decsheikh,Hag18}. We show via simulations that the proposed schemes for both PCs and SCCs provide up to $0.88$ dB gain compared to standard HDD based on iBDD of the component codes for a bit-interleaved coded modulation (BICM) scheme using $256$ quadrature amplitude modulation (QAM). Such gains are predicted to yield up to $33\%$ optical reach improvement compared to the original optical reach of iBDD for BICM with $256$-QAM.

The remainder of the paper is organized as follows. In Section~\ref{Sec:Prel}, some preliminaries and the system model are explained, and the iBDD-CR algorithm is reviewed. The new hybrid decoding algorithm for PCs and SCCs is introduced in Sections~\ref{Sec:BEE-PC} and \ref{Sec:BEE-SCC}, respectively. In Section~\ref{Sec:HandC} we explain the heuristics behind the proposed hybrid decoders and analyze their complexity. Conclusions are drawn in Section~\ref{Sec:Conclusions}. \footnote{\textbf{Notation}: Boldface letters stand for vectors
	and matrices, e.g., $\boldsymbol{x}$  and $\boldsymbol{X}=[x_{i,j}]$, with $x_{i,j}$ representing the element corresponding to the $i$-th row and $j$-th column of $\boldsymbol{X}$. $\boldsymbol{X}_{i,:}$ denotes the $i$-th row of $\boldsymbol{X}$. A Gaussian distribution with mean $\mu$ and variance $\sigma^2$ is denoted by $\mathcal{N}(\mu ,\sigma^2)$. Functions and subscript texts are shown with $\mathsf{serif}$ font. $(\cdot)^\mathsf{T}$ is the matrix transpose operation. $|x|$ and $|\boldsymbol{x}|$ stand for the absolute value of $x$ and the vector with components equal to the absolute value of the component of $\boldsymbol{x}$, resp.} 

 \begin{figure}[t!] \centering 
	\includegraphics[scale=0.8]{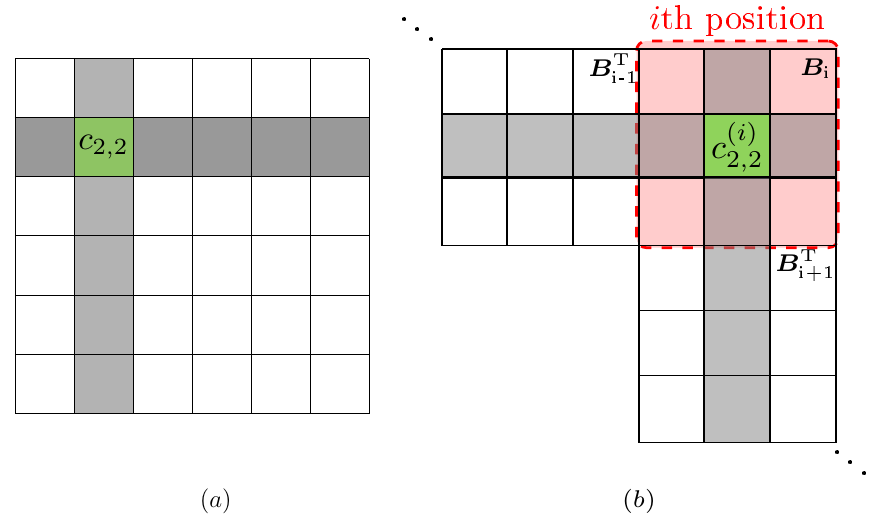}  
	\caption{(a) Schematic of an PC with $\nc=6$. The code bit corresponding to the second row and second column codes is marked in green. (b) Schematic of an SCC with $\nc=6$ containing blocks $\Bmat_{i-1}$, $\Bmat_{i}$, and $\Bmat_{i+1}$. The code bit corresponding to the second row and column codes in $\Bmat_{i}$ is marked in green.}  
	\label{PC_SCC} 
\end{figure}  

\section{Preliminaries}\label{Sec:Prel}

Product-like codes are a family of codes described by a two-dimensional array.\footnote{While higher-dimensional product-like codes exist, the most conventional ones are two-dimensional.} In this paper, we consider binary PCs and SCCs. In the following, we briefly review the structure of PCs and SCCs and also explain the channel model considered in this paper. The last part of this section describes the recently introduced iBDD-CR algorithm.

\subsection{Product and Staircase Codes}

Let $\mathcal{C}$ be a Bose-Chaudhuri-Hocquenghem (BCH) code constructed over the Galois field $\text{GF}(2^v)$ with error correction capability $t$ and shortening parameter $s$. The codeword length of $\mathcal{C}$ is $\nc=2^v-1-s$ and the number of its information bits is $\kc=2^v-vt-1-s$. A PC with $(\nc,\kc)$ component codes is defined as the set of all $\nc \times \nc$ arrays $\cc=[c_{i,j}]$ such that each row and column of $\cc$ is a valid codeword of $\mathcal{C}$. The rate of such PC is $R=\kc^2/\nc^2$. Fig.~\ref{PC_SCC}(a) shows the  code array of a PC code with component code length $n=6$. The code bit corresponding to the second row and second column component codes,  $c_{2,2}$, is highlighted. PCs are conventionally decoded based on BDD of the component codes. BDD corrects all error patterns of Hamming weight up to $t$. If the weight of the error pattern is larger than $t$ and there exists another codeword with Hamming distance less than $t$ to the received codeword, BDD introduces miscorrections. Otherwise, BDD fails, where conventionally it is considered that BDD outputs its input. We refer to iteratively applying BDD on row and column codes as iBDD. 

A SCC comprises the set of all matrices $\Bmat_i$ of size $\nc/2 \times \nc/2$, $i=1,2,\ldots$, such that each row of the matrix $[\Bmat_{i-1}^\rmT,\Bmat_i]$ is a valid codeword in $\mathcal{C}$. Each matrix $\Bmat_i$ contains $(\nc/2)\cdot(\nc-\kc)$ parity bits out of $\nc^2/4$ code bits, hence, the corresponding code rate is $R = 1- 2(\nc-\kc)/\nc$. Fig.~\ref{PC_SCC}(b) shows an schematic of the SCC comprising the blocks $\Bmat_{i-1}$, $\Bmat_{i}$, $\Bmat_{i+1}$, and $c^{(i)}_{2,2}$ as a code bit corresponding to the second row and second column of $\Bmat_{i}$. Note that $\Bmat_0$ is an all-zero matrix which initializes the decoding procedure of the SCC. SCCs are decoded in a windowed-decoding fashion, i.e., each row of $[\Bmat_{i-1}^\rmT,\Bmat_i]$ for staircase blocks within a decoding window is decoded based on BDD \cite{staircase_frank}. We also refer to iteratively applying BDD on the component codes of an SCC within the decoding window as iBDD.

\subsection{System Model}
We consider bit-interleaved coded modulation (BICM)  based on  $M^2$-QAM. In BICM, the code bits are interleaved and then mapped to the constellation points using the binary reflected Gray code (BRGC) mapping. The real and imaginary part of an $M^2$-QAM symbol with $M=2^{m}$ are both selected from the set $\mathcal{X}\triangleq\{(-2^m+1)\cdot \Delta ,...,-\Delta,\Delta,...,(2^m-1)\cdot \Delta\}$, where $\Delta=\sqrt{3/2(M^2-1)}$ normalizes the constellation energy to unity. Due to the symmetry of the $M^2$-QAM constellation, we only consider transmission of the real part of the $M^2$-QAM symbol. 

The AWGN channel output at time instant $i$ corresponding to transmitted symbol $x_{i} \in \mathcal{X}$ is given by
\begin{equation}\label{awgn}
	y_{i}=x_{i}+n_{i},\;\;\;\;\;i=1,2,\cdots,n_\mathsf{ch},
\end{equation}  
where $n_\mathsf{ch}$ is the number of channel uses corresponding to a PC or SCC block, and $n_{i}\sim \mathcal{N}(0,\sigma^2)$. The log likelihood ratio (LLR) of the \AS{$q$th bit level of $y_{i}$ is given as 
\begin{equation}
l_i^q = \sum\limits_{b \in \left\{ {0,1} \right\}} {{{\left( { - 1} \right)}^b}} \ln  {\sum\limits_{a \in {\cal S}_q^b} {{e^{ - \frac{{{{({y_i} - a)}^2}}}{{2{\sigma ^2}}}}}} } ,\;q=1,2,\cdots,m,
\end{equation} 
where $\mathcal{S}_q^0 \subset \mathcal{X}$ and $\mathcal{S}_q^1 \subset \mathcal{X}$ are sets of size $2^{m}$ symbols with $0$ and $1$ as the $q$th bit of the corresponding BRGC label, respectively.} For the special case of the binary input AWGN (bi-AWGN) channel, $m=1$, $\Delta=1$, and $l_i=2y_{i}/{\sigma^2}$.

\begin{figure}[t] \centering 
	\includegraphics[scale=0.72]{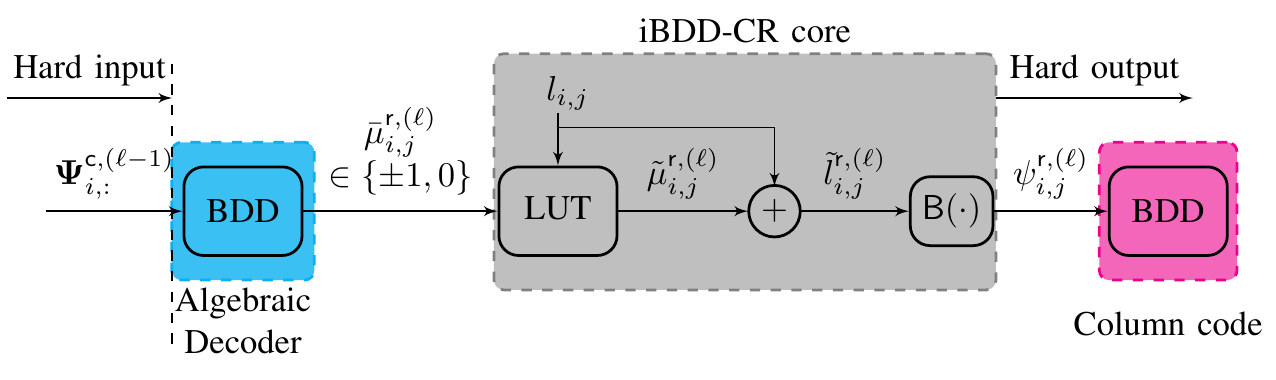}  
	\caption{Schematic of iBDD-CR \cite{optimal_decsheikh} for decision on code bit $c_{i,j}$ at iteration $\ell$ by $i$th row decoding with input $\bm{\Psi}^{\mathsf{c},(\ell-1)}_{i,:}$.}  
	\label{iBDD-CRPC} 
\end{figure}

\begin{figure*}[t] \centering 
	\includegraphics[scale=0.8]{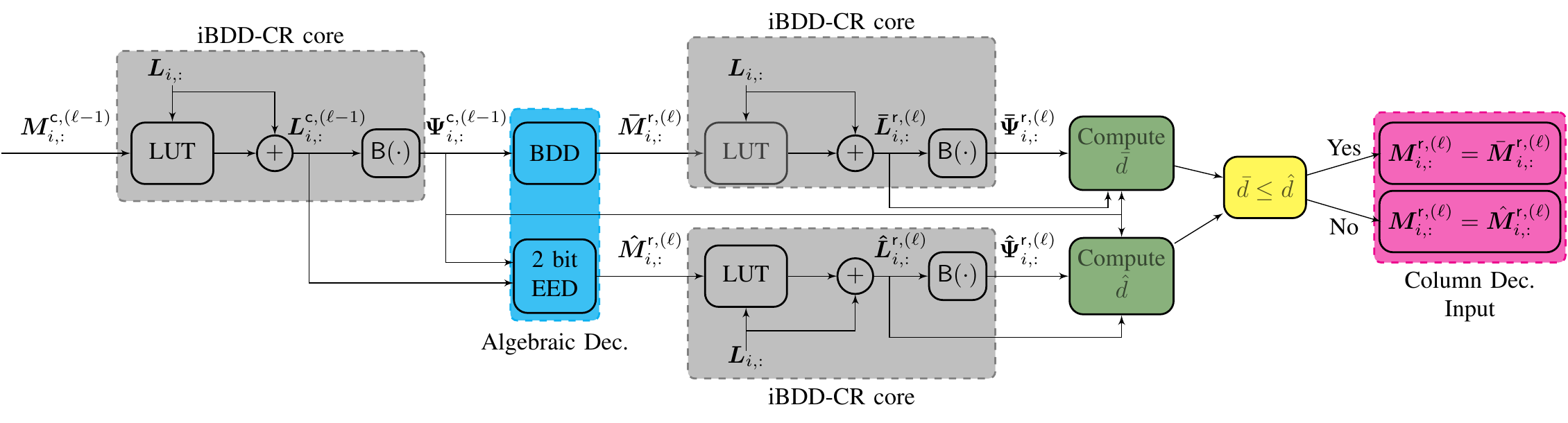}  
	\caption{Schematic of BEE-PC corresponding to decoding of the $i$th row of a PC at iteration $\ell$.}  
	\label{block_BEE-PC} 
\end{figure*} 

\subsection{iBDD with Combined Reliability}\label{RABMP_PC}

We briefly review the architecture of iBDD-CR \cite{optimal_decsheikh}, which is shown in Fig.~\ref{iBDD-CRPC}.

Let us denote by $\bm{\Psi}^{\mathsf{c},(\ell-1)}=[\psi_{i,j}^{\mathsf{c},(\ell-1)}]$ the decoding output of the $\nc$ column codes at iteration $\ell-1$, i.e., $\psi_{i,j}^{\mathsf{c},(\ell-1)}$ corresponds to the hard decision on code bit $c_{i,j}$. The input of the row decoder at iteration $\ell$ is $\bm{\Psi}^{\mathsf{c},(\ell-1)}$. In the following, we  explain the decoding of the $i$th row code at iteration $\ell$, using the input $\bm{\Psi}^{\mathsf{c},(\ell-1)}_{i,:}$. We denote by $\bar\mu_{i,j}^{\mathsf r}$ the output of BDD of the $i$th row code corresponding to code bit $c_{i,j}$. In case of correct decoding or miscorrection, BDD outputs a codeword. In this case  $\bar\mu_{i,j}^{\mathsf r} \in \{ \pm 1 \}$ using the mapping $0\mapsto +1$ and $1\mapsto -1$. Furthermore, in case of BDD failure, we set $\bar\mu_{i,j}^{\mathsf r}=0$.\footnote{Throughout this paper we repeateadly use a ternary alphabe $\{-1,0,+1\}$. We use $\{\pm1\}$ to denote correct decoding or miscorrections using the mapping $0\mapsto +1$ and $1\mapsto -1$, and $0$ to represent a decoding failure.}

The LLR on code bit $c_{i,j}$ after BDD at iteration $\ell$ is given as \cite[Eq.~(9)]{optimal_decsheikh}
\begin{align}\label{LLRoutv1} 
\opt_{i,j}^{\row,(\ell )} = \mut_{i,j}^{\row,(\ell )} + l_{i,j}, 
\end{align}
where $l_{i,j}$ is the channel LLR and $\mut_{i,j}^{\row,(\ell )}$ can be computed using a look-up table (LUT) based on $\bar\mu_{i,j}^{\mathsf r}$ (see \cite[Theorem.~1]{optimal_decsheikh}). Then, a hard decision is made on $\opt_{i,j}^{\row,(\ell )}$,
\begin{equation}\label{eq:BDDchrel_VN}
\psi_{i,j}^{\mathsf{r},(\ell)}=
\mathsf{B}\left(\tilde{l}_{i,j}^{\mathsf r, (\ell)}\right),
\end{equation}
where $\BB(\cdot)$ is the mapping $- 1 \mapsto 1$ and $+ 1 \mapsto 0$. After applying this procedure to all row codes, the matrix
$\boldsymbol{\Psi}^{\mathsf{r},(\ell)}=[\psi_{i,j}^{\mathsf{r},(\ell)}]$ is formed and used as the input for the $n$ column decoders. Column decoding is similar to the row counterpart. The decoding continues by iterating between row and column decoding for a given number of iterations. The additional component of iBDD-CR compared to iBDD is shown as ``iBDD-CR core'' in Fig.~\ref{iBDD-CRPC}. As can be seen, the operations in the iBDD-CR core are based on soft values; however, only (binary) hard messages are exchanged between component decoders. Therefore, the contribution of the messages exchange between component decoders to the overall internal decoder data flow of iBDD-CR is the same as that of iBDD \cite{optimal_decsheikh}.  

\section{Hybrid Decoding of PCs}\label{Sec:BEE-PC}

In this section, we  introduce a novel hybrid decoding algorithm for PCs, which we call binary message passing based on EDD (BEE), as an improved variant of our recently introduced iBDD-CR algorithm.  BEE is then extended to SCCs in Section~\ref{Sec:BEE-SCC}. We will refer to BEE for PCs and SCCs as BEE-PC and BEE-SCC, respectively.

\subsection{Generalized Distance Metric}

BEE exploits the so-called generalized distance (GD) metric, originally introduced in \cite{ForneyGMD}, in order to improve the performance of iBDD-CR. GD is defined as follows. Let $\a=(a_1,a_2,\cdots,a_n)$ be a binary vector with reliability $\boldsymbol{l}=(l_1,l_2,\cdots,l_n)$. The GD between $\a$ and the binary vector $\b=(b_1,b_2,\cdots,b_n)$ is
\begin{align}\label{GMDmetric}
\scalemath{1}{
	\dGMD  \buildrel \Delta \over =  \sum_{\substack{i=1\\a_i=b_i}}^{n}
	{\left( {1 - {\alpha_{i}}} \right)}  + 
	\sum_{\substack{i=1\\a_i\ne b_i}}^{n}
	{\left( {1 + {\alpha_{i}}} \right)}}, 
\end{align}
where $\boldsymbol{\alpha}=(\alpha_1,\alpha_2,\cdots,\alpha_n)$ is the vector of \emph{normalized} reliabilities, i.e., $\alpha_{i}\buildrel \Delta \over = |l_{i}|/\mathop {\max
}\limits_{1 \le j \le n} |l_{j}|$. One can interpret the GD as a soft version of the Hamming distance. Indeed, assuming $\alpha_{i}=1$ for $a_{i} = {b_{i}}$ and $\alpha_{i}=0$ for $a_{i} \ne {b_{i}}$, $\dGMD$ corresponds to the Hamming distance between $\a$ and $\b$.  

\subsection{BEE-PC Algorithm}

Fig.~\ref{block_BEE-PC} shows the block diagram of the proposed BEE algorithm for PCs. Without loss of generality, we explain the BEE-PC decision corresponding to the  decoding of the $i$th row of the PC at iteration $\ell$. BEE-PC encompasses two decoding attempts, each corresponding to a branch in Fig.~\ref{block_BEE-PC}. First, we explain the upper branch of Fig.~\ref{block_BEE-PC}. Let $\boldsymbol{M}^{\mathsf{c},(\ell-1)}_{i,:}$ contain the BDD outcome of the $\nc$ column decoders at iteration $\ell-1$ corresponding to the $i$th row of the PC codeword, with elements in $\{\pm1,0\}$. Similar to the iBDD-CR decoder (see \eqref{LLRoutv1}), using $\boldsymbol{M}^{\mathsf{c},(\ell-1)}_{i,:}$, the  channel LLRs $\boldsymbol{L}_{i,:}$, and the iBDD-CR LUT, the LLR vector $\boldsymbol{L}^{\mathsf{c},(\ell-1)}_{i,:}$ can be computed, on which a hard decision is made according to $\boldsymbol{\Psi}^{\mathsf{c},(\ell-1)}_{i,:}=\BB(\boldsymbol{L}^{\mathsf{c},(\ell-1)}_{i,:})$. Then, BDD is performed with  input $\boldsymbol{\Psi}^{\mathsf{c},(\ell-1)}_{i,:}$, which yields ${\boldsymbol{\bar{M}}}_{i,:}^{\mathsf r,(\ell)}$, with elements in $\{\pm1,0\}$. Using  ${\boldsymbol{\bar{M}}}_{i,:}^{\mathsf r,(\ell)}$, $\boldsymbol{L}_{i,:}$, and the iBDD-CR LUT, the LLR $\bar{\boldsymbol{L}}^{\mathsf{r},(\ell)}_{i,:}$ is then computed and the candidate decision $\bar{\boldsymbol{\Psi}}^{\mathsf{r},(\ell)}_{i,:}$ is formed as $\bar{\boldsymbol{\Psi}}^{\mathsf{r},(\ell)}_{i,:}=\BB(\bar{\boldsymbol{L}}^{\mathsf{r},(\ell)}_{i,:})$. Finally, a \emph{score} of the candidate decision $\bar{\boldsymbol{\Psi}}^{\mathsf{r},(\ell)}_{i,:}$, denoted by $\bar{d}$ is obtained as 
\begin{align}
\label{d1compc}
\scalemath{1}{
	\bar{d} =
	\begin{dcases}
	2\nc, &
	\text{if } \text{BDD fails}\\
	{d}_\mathsf{GD}\left(\bar{\boldsymbol{{\Psi}}}^{\mathsf{r},(\ell)}_{i,:},\boldsymbol{\Psi}^{\mathsf{c},(\ell-1)}_{i,:}\right), &
	\text{otherwise} \\
	\end{dcases}},
\end{align} 
where $2n$ is a constant discussed in Section~\ref{Comments:Heuristics}.

The second branch of Fig.~\ref{block_BEE-PC} serves as another decoding attempt based on EED. The two least reliable bits of $\boldsymbol{\Psi}^{\mathsf{c},(\ell-1)}_{i,:}$ are found based on $|\boldsymbol{L}^{\mathsf{c},(\ell-1)}_{i,:}|$ and then erased. Then, algebraic EED \cite[Sec.~6.6]{LinCos04} is performed on $\boldsymbol{\Psi}^{\mathsf{c},(\ell-1)}_{i,:}$. Let $\hat{\boldsymbol{M}}_{i,:}^{\mathsf r,(\ell)}$ be the outcome of EED with three possible values for its elements, i.e., $\{\pm1\}$ in case of successful (but potentially erroneous) decoding with the mapping according to $0\mapsto +1$ and $1\mapsto -1$, and $0$ if EED fails. Similar to what explained for the first branch, a candidate decision 
is formed as $\hat{\boldsymbol{\Psi}}^{\mathsf{r},(\ell)}_{i,:}=\BB(\hat{\boldsymbol{L}}^{\mathsf{r},(\ell)}_{i,:})$, where the LLR vector $\hat{\boldsymbol{L}}^{\mathsf{r},(\ell)}_{i,:}$ is computed using $\hat{\boldsymbol{M}}^{\mathsf{r},(\ell)}_{i,:}$,  $\boldsymbol{L}_{i,:}$, and the iBDD-CR LUT. Finally, the  score of candidate decision $\hat{\boldsymbol{\Psi}}^{\mathsf{r},(\ell)}_{i,:}$ is computed as 
\begin{align}
\label{d2compc}
\scalemath{1}{
	\hat{d} =
	\begin{dcases}
	2\nc, &
	\text{if } \text{EED fails}\\
	{d}_\mathsf{GD}\left(\hat{\boldsymbol{{\Psi}}}^{\mathsf{r},(\ell)}_{i,:},\boldsymbol{\Psi}^{\mathsf{c},(\ell-1)}_{i,:}\right), &
	\text{otherwise} \\
	\end{dcases}}~.
\end{align} 

Finally, comparing $\bar{d}$ and $\hat{d}$, the candidate codeword with the minimum score is chosen as the BEE-PC decision and the corresponding BDD outcome is as the input for column decoding in the next iteration, i.e.,
\begin{align}
\label{fD_PC}
\scalemath{1}{
	{\boldsymbol{M}}_{i,:}^{\mathsf r,(\ell)} =
	\begin{dcases}
	{\bar{\boldsymbol{{M}}}}_{i,:}^{\mathsf r,(\ell)}, &
	\text{if } \bar{d} \le \hat{d}\\
	{\hat{\boldsymbol{{M}}}}_{i,:}^{\mathsf r,(\ell)}, &
	\text{if } \bar{d} > \hat{d}\\
	\end{dcases}}~.
\end{align} 
After decoding of all rows of the PC at iteration $\ell$, ${\boldsymbol{M}}^{\mathsf r,(\ell)}$ is utilized as the input to the column decoders. To initialize the algorithm, the channel LLRs are employed as the input LLRs of both BDD and EED blocks, i.e., $\boldsymbol{L}^{\mathsf{c},(0)}=\boldsymbol{L}$. In the last iteration $(\ell_\text{max})$, the BEE-PC decoding output corresponds to the branch with lowest score, i.e., the decoding output is $\bar{\boldsymbol{\Psi}}^{\mathsf r,(\ell_\text{max})}$ if $\bar{d} \le  \hat{d}$ and  $\hat{\boldsymbol{\Psi}}^{\mathsf r,(\ell_\text{max})}$ if $\bar{d} > \hat{d}$.

\subsection{Numerical Results}\label{num_PC}

\begin{table*}[t]
\caption{Codes used for simulations}
	\centering
\begin{tabular}{ccccccccccccc}
\makecell{Component\\ code} &
\makecell{component\\ parameters} &
\makecell{component\\ code rate} &
\makecell{PC \\ code rate} &
\makecell{HD Shannon limit  \\ at PC code rate} & 
\makecell{SD Shannon limit \\ at PC code rate} &
\makecell{SCC \\ code rate} &
\makecell{HD Shannon limit \\ at SCC code rate} &
\makecell{SD Shannon limit \\ at SCC code rate} \\
\midrule
$\mathcal{C}_1$ & (256,239,2) & 0.933 & 0.871 & 4.05 (dB) & 2.64 (dB) & 0.867 & 3.99 (dB) & 2.74 (dB) \\
$\mathcal{C}_2$ & (255,231,3) & 0.905 & 0.820 & 3.54 (dB) & 2.23 (dB) & 0.811 & 3.46 (dB) & 2.14 (dB)\\
$\mathcal{C}_3$ & (511,484,3) & 0.947 & 0.897 & 4.36 (dB) & 3.15 (dB) & 0.894 & 4.32 (dB) & 3.11 (dB) \\
\midrule
\end{tabular}
\label{codes}
\end{table*}

\begin{figure}[t] \centering 
	\includegraphics[scale=0.82]{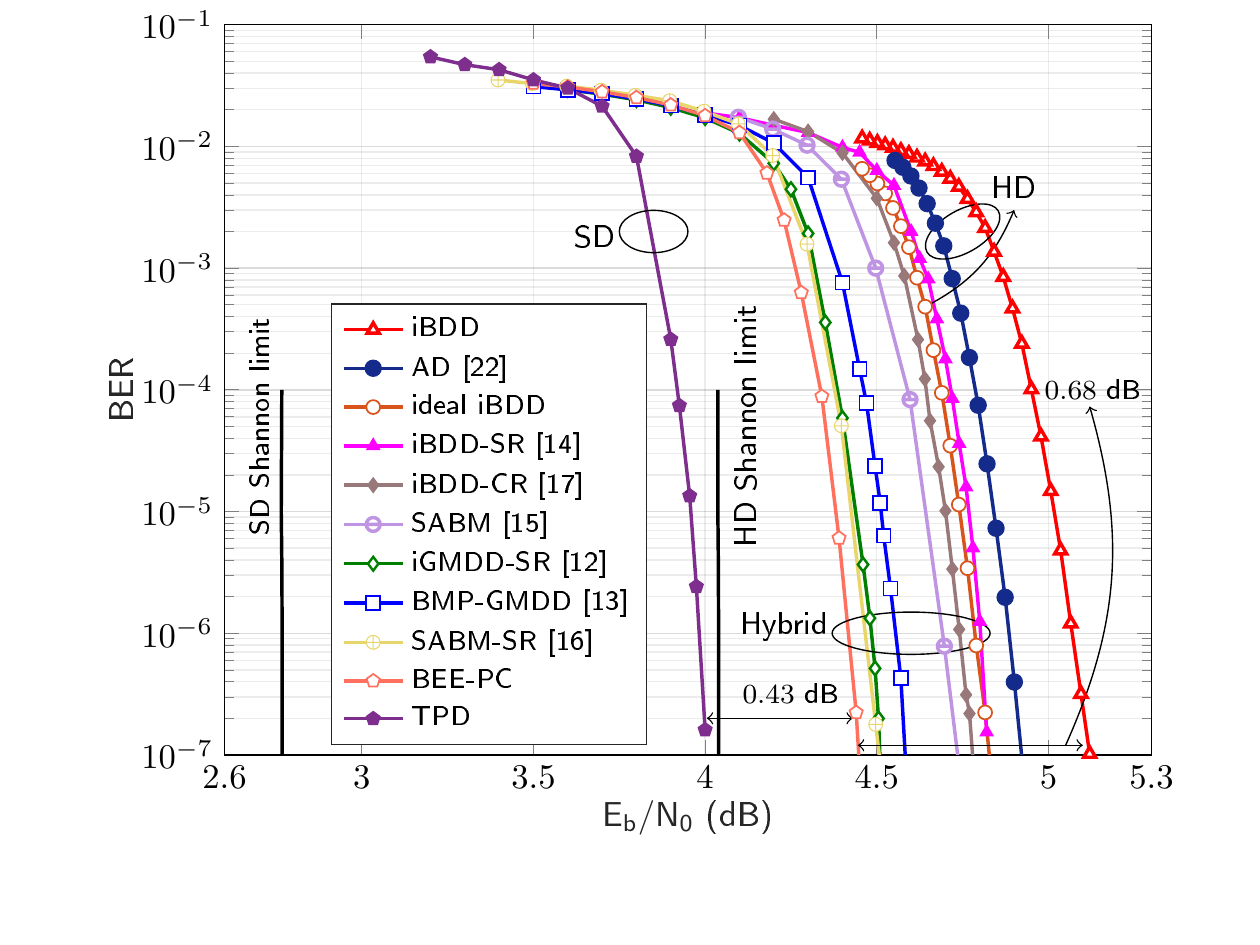}  
	\caption{Performance comparison of different decoders for PC with component code $\mathcal{C}_1$.}  
	\label{simPCbi} 
\end{figure}

\begin{figure*}[t] \centering 
	\includegraphics[scale=0.7]{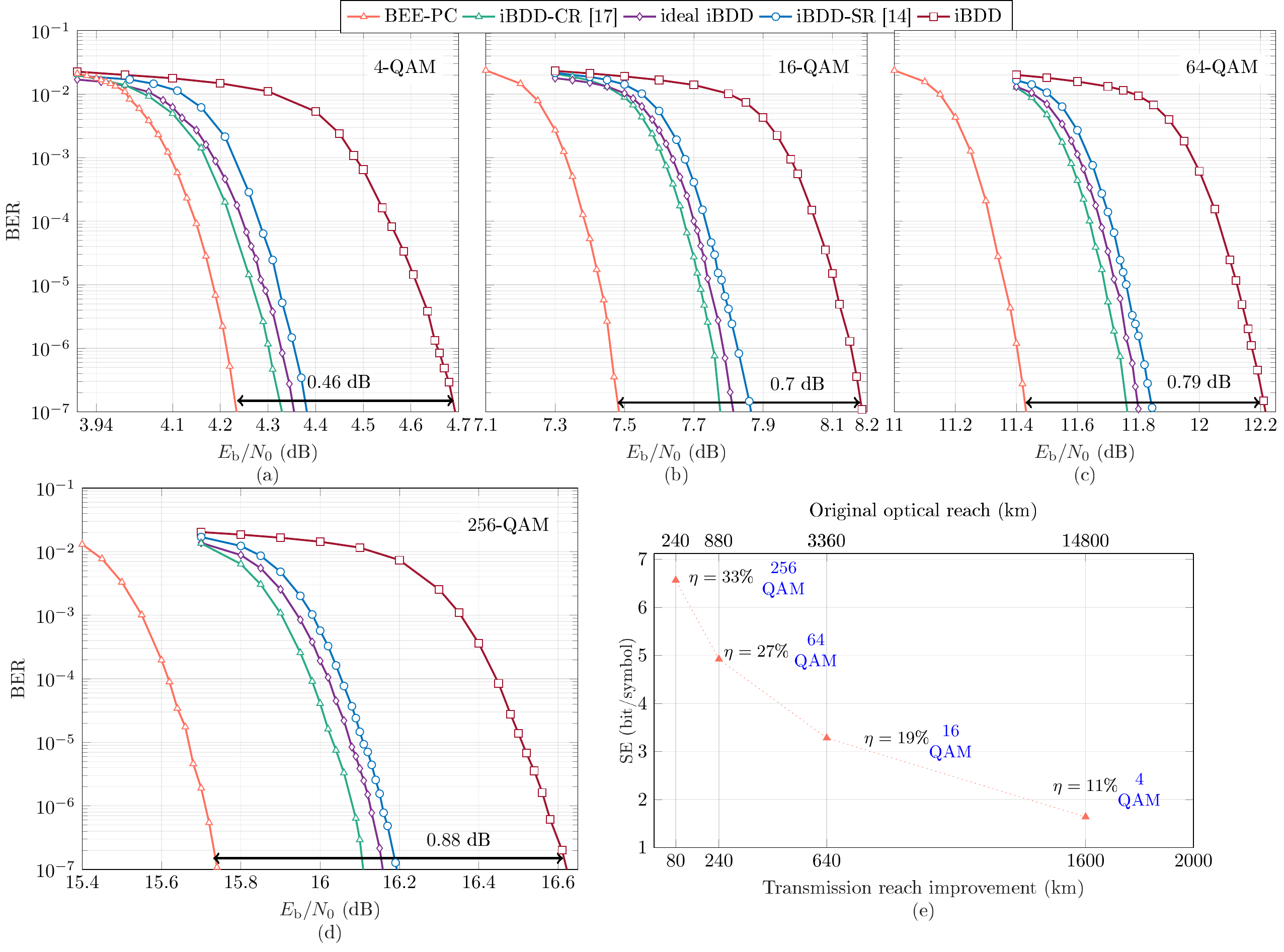}  
	\caption{Performance of iBDD, ideal iBDD, iBDD-SR, iBDD-CR, and BEE-PC for PC with component code $\mathcal{C}_2$ in the BICM with (a) $4$-QAM, (b) $16$-QAM, (c) $64$-QAM, and (d) $256$-QAM modulation. (e) The optical reach improvement of BEE-PC over iBDD as well as the original optical reach of iBDD, corresponding to (a)-(d).}  
	\label{QAM_mod_all_schemes} 
\end{figure*}

We evaluate the performance of BEE-PC. Throughout this paper, we consider an extended BCH (eBCH) code, $\mathcal{C}_1$, and BCH codes $\mathcal{C}_2$ and $\mathcal{C}_3$ as  component codes. The corresponding parameters, PC and SCC rates, and hard decision (HD) and soft decision (SD) Shannon limits  are given in Table~\ref{codes}.\footnote{These component codes are also considered for performance evaluation of the hybrid decoding schemes in \cite{Hag18,She18b,sheikhTCOM19,Yia2019,She19,GabrieleSABMSR2019,optimal_decsheikh}.} 
We  consider $10$ iterations of BEE-PC appended with $2$ iBDD iterations. 
The decision rule  \eqref{eq:BDDchrel_VN} is unable to correct errors with high reliability (code bits in error with high $|{l}_{i,j}|$). This is because  $\mathsf{B}\left(\tilde{l}_{i,j}^{\mathsf r, (\ell)}\right) = \mathsf{B}({l}_{i,j})$,
i.e., the decision on a code bit is overridden by the channel error. Therefore, the appended iBDD iterations (which disregard reliabilities in the decision rule) help the decoder to correct  errors with high reliability (see \cite[Sec.~VI]{sheikhTCOM19}). For the sake of fairness, we evaluate all other algorithms with a total of $12$ iterations.

In Fig.~\ref{simPCbi}, we show the bit error rate (BER) performance of BEE-PC for a PC with component code $\mathcal{C}_1$ and transmission over the bi-AWGN channel. For the sake of comparison, we also depict the performance of iBDD, AD \cite{Hag18}, iBDD-SR \cite{sheikhTCOM19}, SABM \cite{Yia2019}, iBDD-CR \cite{optimal_decsheikh}, iGMDD-SR \cite{She18b}, and SABM-SR \cite{GabrieleSABMSR2019}. We also plot the performance of ideal iBDD which disregards miscorrections using a genie approach, TPD based on the Chase-Pyndiah algorithm \cite{Pyn98}, and the HD and SD Shannon limits. We highlight that iBDD and AD should be compared to the HD capacity. However, all other algorithms should be compared to the SD capacity, as they all exploit channel LLRs in the decoding rule. As it can be seen, BEE-PC outperforms all other algorithms. The performance gain of BEE-PC over conventional iBDD is $0.68$ dB. Remarkably, the gap between BEE-PC and TPD is only $0.43$ dB. Therefore, BEE-PC can close $62\%$ of the gap between (full hard) iBDD and (full soft) TPD. We highlight that the performance of iGMDD-SR, SABM-SR, and BMP-GMDD is close to that of BEE-PC. However, BMP-GMDD requires up to $30\%$ more message exchanging between component decoders than BEE-PC, and both iGMDD-SR and SABM-SR require exchanging soft messages between component decoders, which yields significantly higher decoder data flow than BEE-PC (see Section~\ref{Comments:Complexity}  for a high level complexity discussion of the different algorithms).


\begin{table}[t]	
	\caption{System parameters of a transmission system used in computing the optical reach}
	\centering
	\renewcommand{\arraystretch}{1.2}
	\scalebox{1.2}{	
		\begin{tabular}{c|c}
			\toprule
			\midrule
			Symbol rate: $32$ Gbaud & Channel spacing: $32$ GHz  \\
			No. of Channels: 5 & Roll-off-factor: 0  \\
			$\gamma$: $1.3$ W/km & $D$: 17 ps/nm/km  \\
			$\alpha$: $0.2$ dB/km & $\lambda$: $1550$ nm   \\
			EDFA noise figure: $4.5$ dB & Span length: $80$ km  \\			
			\midrule
			\bottomrule
		\end{tabular}
	}
	\label{Tabcomp1}
\end{table}

In Figs.~\ref{QAM_mod_all_schemes}(a)-(d), the performance of iBDD, ideal iBDD, iBDD-SR, iBDD-CR, and BEE-PC for a PC with component code $\mathcal{C}_2$ are shown for a BICM system (employing random interleaver) with $4$-QAM, $16$-QAM, $64$-QAM, and $256$-QAM, respectively. 
As can be seen, BEE-PC outperforms all other decoders and the gain with respect to iBDD increases using higher order modulation. In particular, the performance gain of BEE-PC over iBDD is $0.46$ dB, $0.7$ dB, $0.79$ dB, and $0.88$ dB for $4$-QAM, $16$-QAM, $64$-QAM, and $256$-QAM, respectively. Fig.~\ref{QAM_mod_all_schemes}(e) shows the spectral efficiency (SE) versus the optical reach improvement of BEE-PC over iBDD as well as the original optical reach of iBDD for  a BICM system employing the same PC as the one  considered in Figs.~\ref{QAM_mod_all_schemes}(a)-(d). In order to predict the transmission reach, we used the enhanced Gaussian noise model\cite{Carena:14} with fiber parameters summarized in Table~\ref{Tabcomp}, evaluated at the optimal launch power. The reach increase for $4$-QAM, $16$-QAM, $64$-QAM, and $256$-QAM are $1600$ km, $640$ km, $240$ km, and $80$ km, respectively. \AS{We also show $\eta$, as the percentage of reach improvement of BEE-PC over iBDD normalized to the original reach of iBDD. The reach improvement increases with increasing modulation order; a $33\%$ reach improvement is attained for $256$-QAM.} 

\section{Hybrid decoding of SCCs}\label{Sec:BEE-SCC}

In this section, we extend iBDD-CR to SCCs by analyzing the decoding behavior using DE for the SC-GLDPC code ensemble, as the  ensemble encompassing SCCs. Then, we propose a novel decoding algorithm for SCCs, similar in spirit to  BEE-PC,
which improves upon iBDD-CR.

\subsection{iBDD-CR decoding of SCCs}

Without loss of generality, we assume that $\Bmat_{i-1}$ and $\Bmat_{i}$ are within the decoding window and we explain the decision of iBDD-CR at iteration $\ell$ corresponding to $c^{(i)}_{j,p}$, which is located in $\Bmat_i$, and also in the $j$-th row of $[\Bmat_{i-1}^\rmT,\Bmat_i]$. Let us assume that $\bm{\Psi}^{(i),(\ell)}=[\psi_{j,p}^{(i),(\ell)}]$, $j=1,2,\cdots,\frac{\nc}{2}$, $p=1,2,\cdots,\nc$ contains the hard decision inputs of iBDD-CR corresponding to $[\Bmat_{i-1}^\rmT,\Bmat_i]$, and $\bm{L}^{(i)}=[l_{j,p}^{(i)}]$ is the corresponding channel LLRs.   

We denote by $\bar\mu_{j,p}^{(i)}\in\{\pm1, 0 \}$ the output of BDD on the $j$th row of $\bm{\Psi}^{(i),(\ell)}$, i.e., $\bm{\Psi}^{(i),(\ell)}_{j,:}$, corresponding to code bit $c^{(i)}_{j,p}$. Furthermore,  let $\tilde{l}_{j,p}^{(i), (\ell)}$ be the LLR of code bit $c^{(i)}_{j,p}$ after BDD at  decoding iteration $\ell$.
The hard decision on code bit $c^{(i)}_{j,p}$ produced by the $j$th row decoder is formed as
\begin{equation}\label{eq:BDDchrel_VN1}
\psi_{j,p}^{(i),(\ell)}=
\mathsf{B}\left(\tilde{l}_{j,p}^{(i), (\ell)}\right).
\end{equation}
After updating $\bm{\Psi}^{(i),(\ell)}$, $\bm{\Psi}^{(i+1),(\ell)}$ is used as the input to the iBDD-CR decoder, which decodes each row of $[\Bmat_{i}^\rmT,\Bmat_{i+1}]$. This procedure continues until all blocks within the decoding window are decoded for a given number of iterations. 

In Fig.~\ref{iBDD-CR}, a schematic of the iBDD-CR decision for the code bit $c^{(i)}_{j,p}$ corresponding to the decoding iteration $\ell$ is shown.  As can be seen from the figure, the core of iBDD-CR is to compute $\tilde{l}_{j,p}^{(i), (\ell)}$. As explained in Section~\ref{RABMP_PC}, $\tilde{l}_{j,p}^{(i),(\ell )} = \mut_{j,p}^{(i),(\ell )} + l_{j,p}$,
where $\mut_{j,p}^{(i),(\ell )}$ can be computed using a LUT (see \cite[Table I \& II]{optimal_decsheikh}). In \cite{optimal_decsheikh} the entries of the LUT are optimized via DE for the GLDPC code ensemble encompassing PCs, and used for implementing the iBDD-CR for PCs. In the following, we extend this DE analysis to the SC-GLDPC code ensemble, which allows to find the LUT for SCCs and implement  iBDD-CR.

\subsection{DE Analysis of iBDD-CR for SC-GLDPC Ensembles}\label{DE_analysis}

\begin{figure}[t] \centering 
	\includegraphics[scale=0.72]{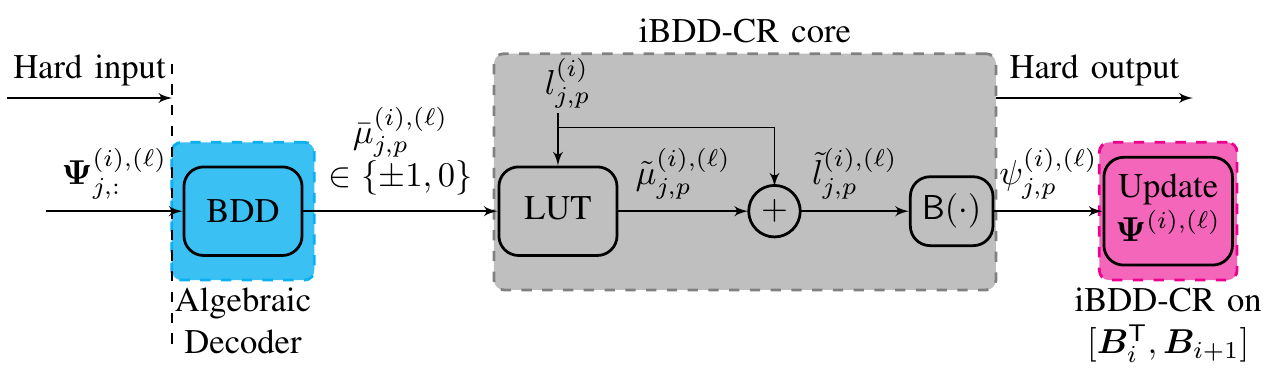}  
	\caption{Schematic of iBDD-CR for decision on code bit $c^{(i)}_{j,p}$ at iteration $\ell$.}  
	\label{iBDD-CR} 
\end{figure}

 \begin{figure}[t!] \centering 
	\includegraphics[scale=1]{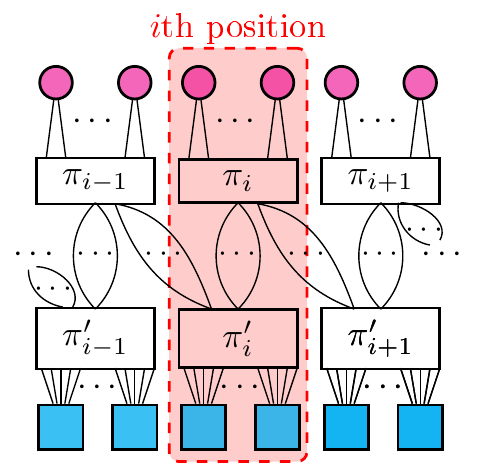}  
	\caption{The SC-GLDPC ensemble with $\nc^{2}/4$ degree-2 VNs and $\nc/2$ CNs of degree-$\nc$. The $i$th spatial position is shaded in red. $\pi_{i}$ and $\pi^{\prime}_{i}$ are random interleavers corresponding to VNs and CNs, resp., at $i$th spatial position.}  
	\label{STC_graph} 
\end{figure}   

Fig.~\ref{STC_graph} shows the Tanner graph of a SC-GLDPC ensemble with $\nc^{2}/4$ degree-2 variable nodes (VNs) and $\nc/2$ constraint nodes (CNs) of degree $\nc$ \AS{at each spatial position}. 

The coupling memory of this SC-GLDPC ensemble is $2$. Therefore, the VNs at spatial position $i$ are randomly connected to CNs at spatial position $i$ and $i+1$. This randomness is represented in Fig.~\ref{STC_graph} by the edge interleavers $\pi_{i}$. Furthermore, CNs at spatial position $i$ are randomly  connected 
(via  the edge interleavers $\pi^{\prime}_{i}$  in Fig.~\ref{STC_graph}) to VNs at spatial position $i-1$ and $i$. Comparing Fig.~\ref{PC_SCC}(b) with Fig.~\ref{STC_graph}, one can  infer that SCCs can be constructed as a particular instance of a SC-GLDPC ensemble, where $\nc/2$ BCH component codes correspond to CNs and the $\nc^{2}/4$ code bits of $\Bmat_i$ correspond to VNs at spatial position $i$. 
We highlight that SCCs are deterministic codes (see \cite{Haeger2017tit}), i.e., the connections between VNs and CNs are determined by the SCC structure, which is not random. The reason for analyzing the SC-GLDPC code ensemble instead of the Tanner graph of SCCs is that the randomization in the SC-GLDPC ensemble significantly simplifies the DE analysis. Assuming extrinsic message passing \cite[Sec.~II--B]{JianPfister2017}, BICM channel mixing \cite[Sec.~IV--A]{Xiechanmix},\footnote{These assumptions are extensively discussed in \cite[Sec.~IV]{optimal_decsheikh}. We remark that these assumptions are only considered for the sake of DE analysis, yielding the LUT required for implementing the iBDD-CR for SCCs. The LUT obtained via DE is then used for the simulation of SCCs. However, for the simulation results in Section~\ref{Sec:BEE-SCC:NumResults}, we consider standard BICM with intrinsic message passing decoding of SCC. No channel mixing and no channel adapters are used in the numerical simulations.} and employing channel adapters in BICM \cite{Hou2003}, in what follows we generalize the DE analysis of iBDD-CR originally derived for the GLDPC code ensemble in \cite[Sec.~IV]{optimal_decsheikh}, to the SC-GLDPC code ensemble.

Let us assume BCH codes as the CNs. In the DE analysis of iBDD-CR for the GLDPC code ensemble \cite[Sec.~IV]{optimal_decsheikh}, it is shown that for a one-dimensional modulation of order $M$ and noise \AS{standard deviation $\sigma$}, the output message error probability of the iBDD-CR at CNs and iteration $\ell$ is given by 
\begin{equation}\label{DE_RABMP}
\mep_{\mathsf{out}}^{(\ell)}= \mathsf{g}\left(\mep_{\mathsf{in}}^{(\ell)},\sigma,M\right), 
\end{equation}
where $\mathsf{g}(\cdot)$ is defined by \cite[Eq.~(13)]{optimal_decsheikh} 
and $\mep_{\mathsf{in}}^{(\ell)}$ denotes the input message error probability of CNs. Note that $\mep_{\mathsf{in}}^{(\ell)}=p_\text{ch}$ initializes the DE, with $p_\text{ch}$ defined as the channel
output error probability yielded by applying hard detection on the channel LLRs. The main difference between the Tanner graph of GLDPC code ensembles and SC-GLDPC code ensembles is the existence of coupling memory in the Tanner graph of the latter. To incorporate the effect of coupling in the DE analysis, we need to track the message error probability corresponding to each spatial position. As we are interested in iBDD-CR for SCCs, in what follows we consider SC-GLDPC code ensembles with coupling memory $2$ (see Fig.~\ref{STC_graph}). 

We denote by $\mep^{(i),(\ell )}_{\mathsf{CN}}$ the average bit error probability from CNs at spatial position $i$ to connected VNs at positions $i$ and $i-1$. Furthermore,  We denote by $\mep^{(i),(\ell)}$ the average bit error probability from VNs at spatial position $i$ to the connected CNs at spatial positions $i$ and $i+1$. $\mep^{(i),(\ell )}_{\mathsf{CN}}$ and $\mep^{(i),(\ell )}$ can be calculated as
\AS{\begin{align}\label{xc} 
\mep^{(i),(\ell )}_{\mathsf{CN}} &= \frac{{\mep^{(i-1),(\ell )} + \mep^{(i),(\ell )}}}{2}, \\ \label{xc2}\mep^{(i),(\ell+1 )} &= \frac{\mathsf{g}\left(\mep^{(i),(\ell )}_{\mathsf{CN}}, \sigma, M\right)+\mathsf{g}\left(\mep^{(i+1),(\ell )}_{\mathsf{CN}}, \sigma, M\right)}{2},                     
\end{align}}
where \eqref{DE_RABMP} is employed to compute \eqref{xc2}. 

Recall that window decoding is usually employed for decoding of SCCs. To account for the effect of window decoding in the DE analysis, we assume that messages are only exchanged between VNs and CNs within the decoding window.  Concretely, let $\mathcal{W}$ be the set of spatial positions within the decoding window and $\mep^{(i),(\ell)}_\mathcal{W}$  the average bit error probability from VNs at spatial position $i$ within the decoding window. Then, $\mep^{(i),(\ell)}_\mathcal{W}=\mep^{(i),(\ell)}$ if $i \in {{\cal W}}$, otherwise $\mep^{(i),(\ell)}_\mathcal{W}=0$.

Employing $\mep^{(i),(\ell)}_\mathcal{W}$ in \eqref{xc}--\eqref{xc2} yields the DE recursion for the SC-GLDPC code ensemble at position $i$ and iteration $\ell+1$, which is given as
\AS{\begin{align}\label{DErec} 
\scalemath{0.88}{\mep^{(i),(\ell+1)}_\mathcal{W} = \frac{{\mathsf{g}\left(\frac{\mep^{(i-1),(\ell)}_\mathcal{W}+\mep^{(i),(\ell)}_\mathcal{W}}{2},  \sigma, M\right) + \mathsf{g}\left(\frac{\mep^{(i),(\ell)}_\mathcal{W}+\mep^{(i+1),(\ell)}_\mathcal{W}}{2}, \sigma, M\right)}}{2}.}                
\end{align}} 
As shown in \cite{optimal_decsheikh}, the computation of $\mathsf{g}(\cdot)$ for a GLDPC code ensemble results in a LUT for the values of $\mut_{i,j}^{\row, (\ell )}$ (see Fig.~\ref{iBDD-CRPC}). Similarly, the computation of \eqref{DErec} for a SC-GLDPC code ensemble yields a LUT for the values of $\mut_{j,p}^{(i),(\ell )}$  corresponding to spatial position $i$ (see Fig.~\ref{iBDD-CR}). Due to this similarity and for the sake of compactness of the paper, we refer to employ \eqref{DErec} in \cite[Proposition 1]{optimal_decsheikh} for computing the entries of the LUTs. 

\subsection{BEE-SCC Algorithm }\label{BEE-SCC_sec}

\begin{figure*}[t] \centering 
	\includegraphics[scale=1.5]{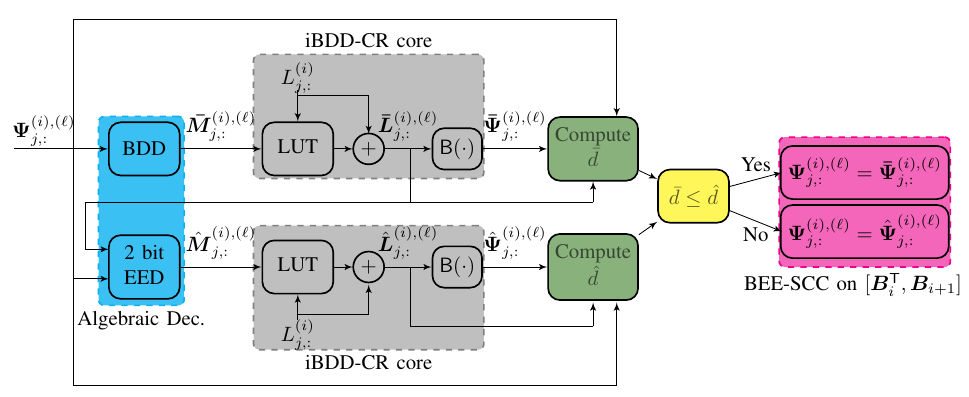}  
	\caption{The schematic of BEE-SCC corresponding to the decision on $j$th row of $[\Bmat_{i-1}^\rmT,\Bmat_i]$ at iteration $\ell$.}  
	\vspace{-3ex}
	\label{BEE-SCC} 
\end{figure*} 

In this section, a novel decoding algorithm  is developed for SCCs.  We refer to this new algorithm as BEE-SCC. Similar to BEE-PC, BEE-SCC is inspired by iBDD-CR and utilizes the GD metric \eqref{GMDmetric}. Fig.~\ref{BEE-SCC} shows the schematic of BEE-SCC in making a decision for the $j$the row of $[\Bmat_{i-1}^\rmT,\Bmat_i]$. Similar to Section~\ref{Sec:BEE-SCC}, we denote by $\bm{\Psi}^{(i),(\ell)}=[\psi_{j,p}^{(i),(\ell)}]$ the hard decision input of BEE-SCC corresponding to $[\Bmat_{i-1}^\rmT,\Bmat_i]$. 

We start by explaining the upper branch of Fig.~\ref{BEE-SCC}, which has some similarities to iBDD-CR. Following the iBDD-CR algorithm with input $\bm{\Psi}^{(i),(\ell)}_{j,:}$, the candidate decision $\bar{\bm{\Psi}}^{(i),(\ell)}_{j,:}$ is computed as 
\begin{equation}\label{Can22}
\bar{\bm{\Psi}}^{(i),(\ell)}_{j,:}=
\mathsf{B}(\bar{\bm{L}}^{(i),(\ell)}_{j,:}) , 
\end{equation}
where $\mathsf{B}(\cdot)$ applies on each component of $\bar{\bm{L}}^{(i),(\ell)}_{j,:}$. Then, by employing \eqref{GMDmetric}, $\bar{{d}}$ defined as the score of candidate decision $\bar{\bm{\Psi}}^{(i),(\ell)}_{j,:}$ is computed as
\begin{align}
\label{d1com}
\scalemath{1}{
\bar{{d}} =
\begin{dcases}
2n, &
\text{if } \text{BDD fails}\\
{d}_\mathsf{GD}(\bm{\bar{\Psi}}^{(i),(\ell)}_{j,:},\bm{\Psi}^{(i),(\ell)}_{j,:}), &
\text{otherwise} \\
\end{dcases}}.
\end{align}

The lower branch of Fig.~\ref{BEE-SCC} serves as a second decoding attempt. In particular, the $2$ least reliable bits of ${\bm{\Psi}}^{(i),(\ell)}_{j,:}$ according to reliability vector $|\bar{\bm{L}}^{(i),(\ell)}_{j,:}|$ are erased and passed to the algebraic EED \cite[Sec.~6.6]{LinCos04}. We denote by $\hat{\bm{M}}^{(i),(\ell)}_{j,:}$ the output of EED with three possible values for the components. Using the same LUT iBDD-CR utilizes, the LLR $\hat{\bm{L}}^{(i),(\ell)}_{j,:}$ is computed. Then, the candidate decision $\hat{\bm{\Psi}}^{(i),(\ell)}_{j,:}$ is formed as
\begin{equation}\label{Can2}
\hat{\bm{\Psi}}^{(i),(\ell)}_{j,:}=
\mathsf{B}(\hat{\bm{L}}^{(i),(\ell)}_{j,:}) , 
\end{equation}     
where $\mathsf{B}(\cdot)$ applies on each component of $\hat{\bm{L}}^{(i),(\ell)}_{j,:}$. Afterwards, $\hat{d}$ as the score of candidate decision $\hat{\bm{\Psi}}^{(i),(\ell)}_{j,:}$ is computed as
\begin{align}
\label{d2comp}
\scalemath{1}{
\hat{d} =
\begin{dcases}
2n, &
\text{if } \text{EED fails}\\
{d}_\mathsf{GD}(\hat{\bm{{\Psi}}}^{(i),(\ell)}_{j,:},\bm{\Psi}^{(i),(\ell)}_{j,:}), &
\text{otherwise} \\
\end{dcases}}.
\end{align}
Finally, the candidate codeword with the minimum score is selected and updates ${\bm{\Psi}}^{(i),(\ell)}_{j,:}$, i.e.,
\begin{align}
\label{fD_SCC}
\scalemath{1}{
	{\bm{\Psi}}^{(i),(\ell)}_{j,:} =
	\begin{dcases}
	{\bar{\bm{\Psi}}}^{(i),(\ell)}_{j,:} &
	\text{if } \bar{d} \le \hat{d}\\
	{\hat{\bm{\Psi}}}^{(i),(\ell)}_{j,:} &
	\text{otherwise} \\
	\end{dcases}}.
\end{align} 

\subsection{Numerical Results}\label{Sec:BEE-SCC:NumResults}
In this section, we evaluate the performance of iBDD-CR and BEE-SCC. We consider a window decoder with a size of $7$ staircase blocks and a maximum of $10$ iterations, appended with $2$ iBDD iterations (see Section~\ref{num_PC} for discussion of appended iBDD iterations). Also, we consider SCCs with even length component codes, hence, when necessary, one bit shortening is performed to have  even component code length. 

\begin{figure}[t] \centering 
	\includegraphics[scale=0.79]{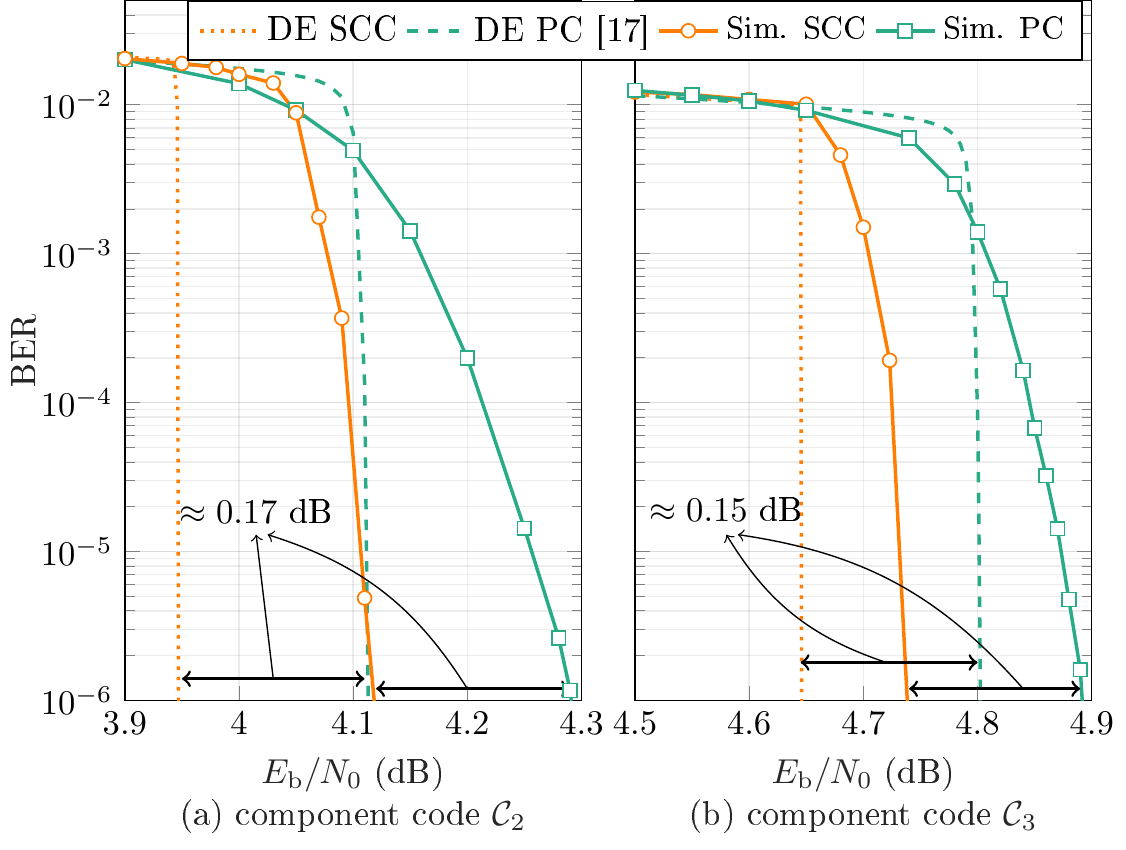}  
	\caption{(a) Comparison between the simulation results of iBDD-CR for SCC and PC as well as the DE analysis of iBDD-CR for SC-GLDPC and GLDPC ensembles for (a) component code $\mathcal{C}_2$ (b) component code $\mathcal{C}_3$.}  
	\vspace{-2ex}
	\label{DE_iBDD-CR} 
\end{figure}

In order to evaluate the derived DE of iBDD-CR for the SC-GLDPC code ensemble (see Section~\ref{DE_analysis}), in Fig.~\ref{DE_iBDD-CR}, we compare the DE results with the performance of SCCs with component codes $\mathcal{C}_2$ and $\mathcal{C}_3$  for transmission over the bi-AWGN channel. For the sake of comparison, we also show the performance of  PCs with the same component codes and the DE evolution curves for the corresponding GLDPC code ensemble \cite[Fig.~4]{optimal_decsheikh}. As it can be seen, DE can predict the performance of SCCs with good accuracy and the gap between simulation results and DE is reduced by increasing the code block length (c.f. the gap between dotted curves with solid curves of Fig.~\ref{DE_iBDD-CR}(a) and Fig.~\ref{DE_iBDD-CR}(b)). Furthermore, the spatial-coupling gain of SCCs over the corresponding PCs is also well-predicted by the DE analysis. Therefore, the derived DE for iBDD-CR in Section~\ref{DE_analysis} can also be used for the parameter optimization of SCCs, similar to the approach taken in \cite{Sheikhecoc2019} for the parameter optimization of PCs. 

\begin{figure}[t] \centering 
	\includegraphics[scale=0.79]{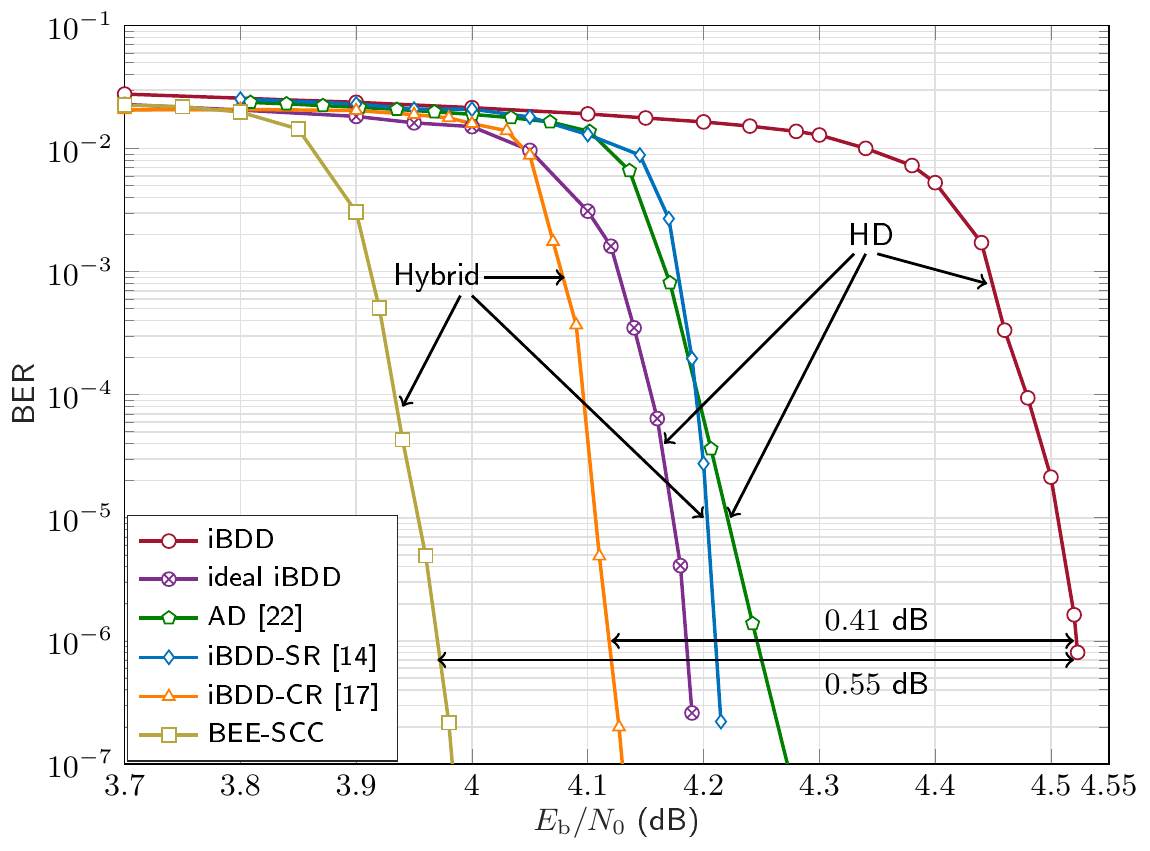}  
	\caption{Performance comparison of iBDD, AD, iBDD-SR, iBDD-CR, and BEE-SCC for staircase code with component code $\mathcal{C}_2$.}  
	\label{sim1} 
\end{figure}

\begin{figure}[t] \centering 
	\includegraphics[scale=0.77]{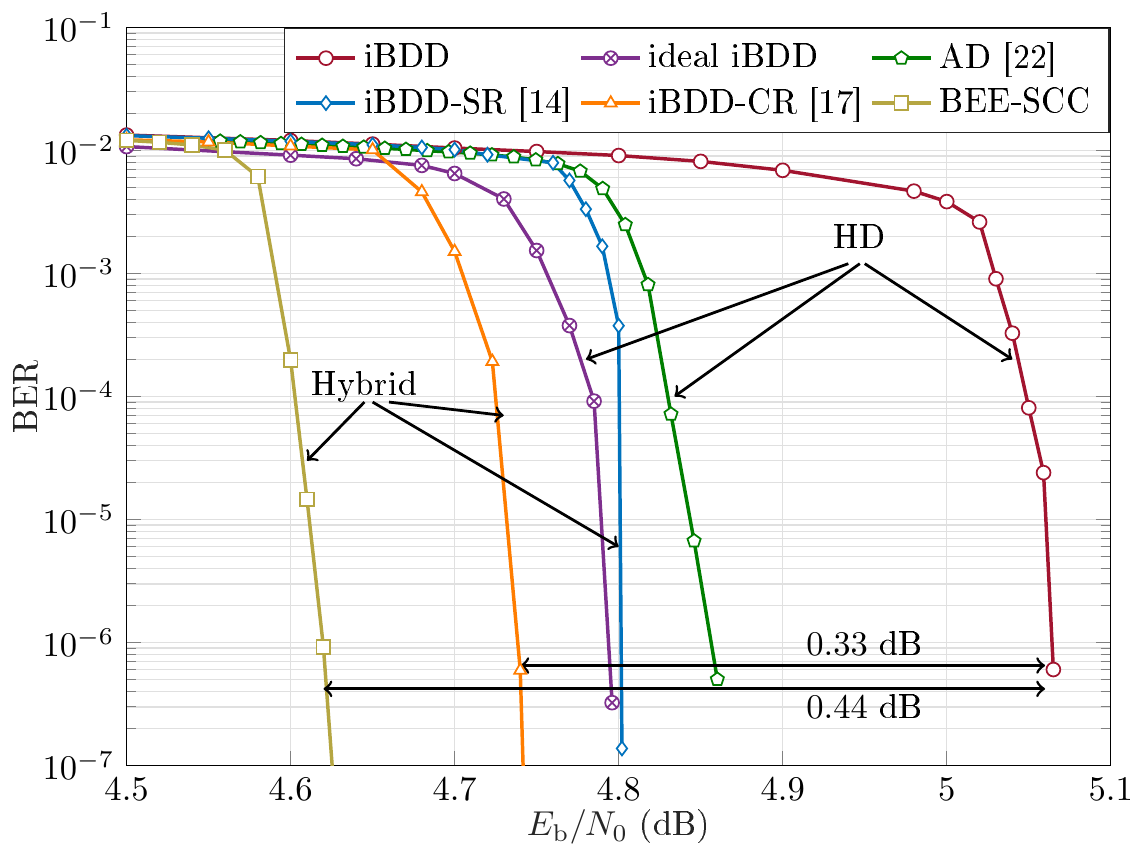}  
	\caption{Performance comparison of iBDD, AD, iBDD-SR, iBDD-CR, and BEE-SCC for SCC with component code $\mathcal{C}_3$.}  
	\label{sim2} 
\end{figure}

In Figs.~\ref{sim1} and \ref{sim2}, the performance of iBDD-CR and BEE-SCC are compared with that of iBDD, ideal iBDD, AD, and iBDD-SR for SCCs with  component codes $\mathcal{C}_2$ and $\mathcal{C}_3$ and transmission over the bi-AWGN channel. One can see that BEE-SCC and iBDD-CR outperform all other algorithms. In particular the gain of iBDD-CR and BEE-SCC over iBDD is $0.41$ dB and $0.55$ dB, respectively, for $\mathcal{C}_2$, and  $0.33$ dB and $0.44$ dB for $\mathcal{C}_3$. 


\begin{figure}[t] \centering 
	\includegraphics[scale=0.79]{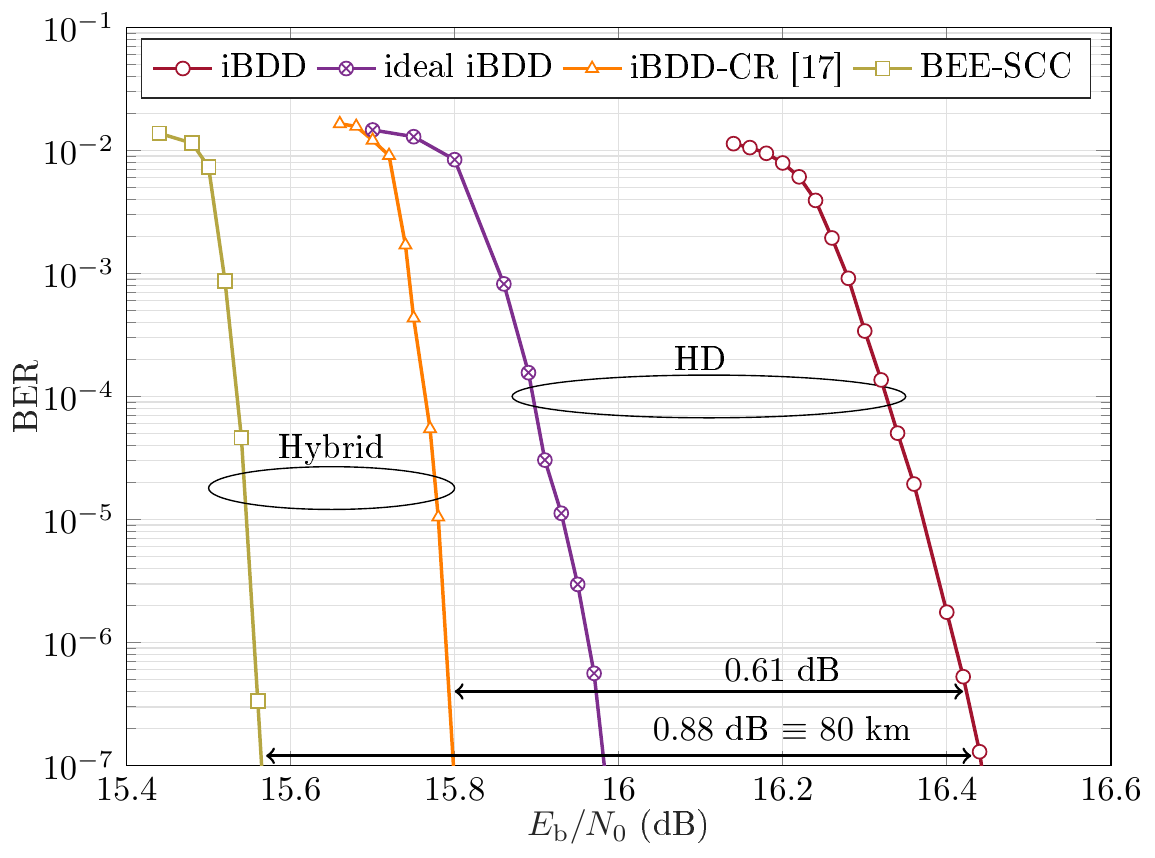}  
	\caption{Performance comparison of iBDD, ideal iBDD, iBDD-CR, and BEE-SCC for a BICM system with $256$-QAM and SCC with component code $\mathcal{C}_2$.}  
	\label{QAM} 
\end{figure}  

In Fig.~\ref{QAM}, we consider a BICM system (employing random interleaver) with $256$-QAM and $\mathcal{C}_2$ as component code, and compare  the performance of iBDD, ideal iBDD, iBDD-CR, and BEE-SCC. For a BER of $10^{-7}$, the performance gain of  iBDD-CR and BEE-SCC over iBDD is $0.61$ dB and $0.88$ dB, respectively. Comparing Fig.~\ref{QAM} with Fig.~\ref{sim1} reveals that the performance improvement of the proposed schemes over iBDD increases by employing higher order modulations. Finally, by employing the GN model and fiber parameters given in Table~\ref{Tabcomp1}, \AS{the gain for BEE-SCC compared to iBDD translates into  $80$ km reach enhancement of BEE-SCC compared to iBDD, corresponding to $\eta=33\%$.}

Comparing Figs.~\ref{simPCbi},~\ref{QAM_mod_all_schemes},~\ref{sim1}, and \ref{sim2}, one can see that the performance gain of BEE-PC and BEE-SCC over iBDD depends on the component codes. In particular, for a given component code length $n$, reducing $t$ yields higher gains. Furthermore, for a given component code error correcting capability $t$, reducing $n$ also increases the performance gain. This is due to fact that the probability of component decoding miscorrections and failures increases in such cases, hence, BEE-PC and BEE-SCC (which deal with recovering miscorrections and failures) can improve more the performance of (standard) iBDD.

\section{BEE-PC and BEE-SCC: Heuristics and Complexity}\label{Sec:HandC}

In this section, our goal is to shed some light on the insights and heuristics at the basis of BEE-PC and BEE-SCC algorithms as well as to discuss the algorithmic complexity of the proposed algorithms.

\subsection{Heuristics}\label{Comments:Heuristics}

We first explain the similarities of  BEE-PC and BEE-SCC. Both algorithms employs an erasure attempt to improve the performance of iBDD-CR. For a component code containing $e$ errors and $s$ erasures, the EED is successful if $2e+s \le {d}_\mathsf{min} -1$,	where ${d}_\mathsf{min}$ is the minimum Hamming distance of the component code \cite[Sec.~6]{LinCos04}. The first attempt of BEE-SCC (corresponding to the upper branches in Fig.~\ref{iBDD-CR} and Fig.~\ref{BEE-SCC}) considers $s=0$, therefore it is capable of correcting up to $\left\lfloor {(d_\mathsf{min}}- 1)/2 \right\rfloor$ errors, i.e., $t$ errors. However, the second attempt (corresponding to the lower branch in Fig.~\ref{iBDD-CR} and Fig.~\ref{BEE-SCC}) considers $s=2$, hence it is capable of correcting $\left\lfloor (d_\mathsf{min}- 3)/2 \right\rfloor$ errors and $2$ erasures. One can easily check that
$\left\lfloor (d_\mathsf{min}- 3)/2 \right\rfloor <  \left\lfloor (d_\mathsf{min} - 1)/2 \right\rfloor$ and $\left\lfloor (d_\mathsf{min} - 3)/2 \right\rfloor +2  \ge  \left\lfloor d_\mathsf{min} - 1/2 \right\rfloor$, therefore, the second attempt can correct more errors if the least reliable bits corresponds to error bits. 

 \newcommand{\tablehighlight}{}
\begin{table*}[t]	
	\caption{Algorithmic-level complexity comparison between different hybrid decoding schemes for PCs and SCCs \cite{Hag18,She18b,sheikhTCOM19,Yia2019,She19,GabrieleSABMSR2019,optimal_decsheikh} and the proposed BEE-PC and BEE-SCC. The memory requirements specify the main type of memory required and a brief explanation.}
	\centering
	\setlength{\tabcolsep}{2pt}
	\renewcommand{\arraystretch}{1}
	\scalebox{0.87}{	
		\begin{tabular}{ccccccc}
			\toprule
			\makecell{\tablehighlight{Decoding}\\ \tablehighlight{Algorithm}} &
			\makecell{\tablehighlight{Computing}\\ \tablehighlight{LLRs}} &
			\makecell{Required \\ sorting Alg.} & 
			\makecell{Code/System \\ type} & \makecell{Contribution of message  \\ exchanges to the decoder data flow} &
			\makecell{\tablehighlight{Memory requirements}}
			& \\ 
			\midrule
			AD \cite{Hag18} & no & no & \makecell{PC, SCC,  \\ bi-AWGN} & \makecell{hard messages \\ similar to iBDD} & \makecell{Dynamic: Storing the location of decoding \\ conflicts and realizing the proposed backtracking algorithm} & \\
			\midrule
			iGMDD-SR \cite{She18b}  & yes & yes & \makecell{PC, bi-AWGN} & soft messages &\makecell{Static: Storing channel LLRs \\ 
				Dynamic: Updating channel LLRs} &  \\
			\midrule
			BMP-GMDD \cite{She19}  & yes &  yes & \makecell{PC, bi-AWGN} & \makecell{hard messages with $8\%$-$33\%$ \\ higher data flow than iBDD} & \makecell{Static: Storing channel LLRs } & 
			\\		
			\midrule
			iBDD-SR \cite{sheikhTCOM19} & yes & no  & \makecell{PC, SCC,  \\ CM} & \makecell{hard messages \\ similar to iBDD} & \makecell{Static: Storing channel LLRs} & \\
			\midrule
			SABM \cite{Yia2019}  & yes & yes & \makecell{PC, SCC,  \\ CM} & \makecell{hard messages \\ similar to iBDD}  & \makecell{Static: Storing the position of highly reliable bits \\ and $d_\mathsf{min}-1-t$ highly unreliable bits for each row and column of PCs \\ and each row of the last staircase block within the decoding window for SCCs} & \\
			\midrule
			SABM-SR \cite{GabrieleSABMSR2019}  & yes & yes & \makecell{PC, CM} &  soft messages  & \makecell{Static: Storing channel LLRs \\ 
				Dynamic: Updating channel LLRs}  &  \\
			\midrule
			iBDD-CR \cite{optimal_decsheikh} & yes & no  & \makecell{PC, SCC,  \\ CM} & \makecell{hard messages \\ similar to iBDD} & \makecell{Static: Storing channel LLRs} &   \\
			\midrule
			\midrule
			BEE-PC  & yes & yes  & \makecell{PC, CM} & \makecell{hard messages with $0.2\%$-$0.5\%$ \\ higher data flow than iBDD} & \makecell{Static: Storing channel LLRs }
			& 
			\\	
			\midrule
			BEE-SCC  & yes & yes  & \makecell{SCC, CM}  & \makecell{hard messages \\ similar to iBDD} & \makecell{Static: Storing channel LLRs } & \\	 					
			\bottomrule
		\end{tabular}
	}
	\label{Tabcomp}
\end{table*} 

For a component code of length $n$, one can show that \eqref{GMDmetric} is upper bounded by $2n$, hence the definition of $\bar{d}$ and $\hat{d}$ for both BEE-PC and BEE-SCC (see \eqref{d1compc}, \eqref{d2compc}, \eqref{d1com}, and \eqref{d2comp}) ensures that the candidate codeword is selected from the branch without a decoding failure. 

We also highlight that both BEE-PC and BEE-SCC employ the same LUT for both decoding attempts. As shown in \cite[Appendix A]{optimal_decsheikh}, the entries of the LUT are derived based on analyzing the behavior of BDD. Following the DE steps in \cite{optimal_decsheikh}, we found that employing EED should in principle yield a new LUT compared to that of iBDD-CR. Unfortunately, finding the entries of the optimal  LUTs for BEE-PC and BEE-SCC requires the computation of some probabilities which seems to be intractable. Therefore, we pragmatically resorted to the same LUT as given by the DE analysis for iBDD-CR. In this sense, the mapping given by the LUTs for the second branch of BEE-PC and BEE-SCC is heuristic.

The main difference between BEE-PC and BEE-SCC is that each algorithm exploits different soft information values for the erasure attempt. In order to efficiently perform the erasure attempt one needs a reliability measure. In BEE-PC, $|\bm{L}^{\mathsf{c},(\ell-1)}_{i,:}|$ is used to find the two least reliable bits of $\bm{\Psi}^{\mathsf{c},(\ell-1)}_{i,:}$(see Fig.~\ref{block_BEE-PC}). If we use the architecture of iBDD-CR to implement BEE-PC, we have to exchange soft values  $\bm{L}^{\mathsf{c},(\ell-1)}_{i,:}$ between component decoders, yielding a significantly higher internal decoder data flow compared to iBDD-CR. To avoid this, BEE-PC modifies the iBDD-CR architecture such that the  BDD/EED decisions, i.e., $\boldsymbol{M}^{\mathsf{c}, (\ell-1)}_{i,:}$ are exchanged (c.f. Fig.~\ref{iBDD-CRPC} and the first branch of Fig.~\ref{block_BEE-PC}). With this modification,  both (hard) decisions $\bm{\Psi}^{\mathsf{c},(\ell-1)}_{i,:}$ and LLRs $\boldsymbol{L}^{\mathsf{c},(\ell-1)}_{i,:}$ used in BEE-PC can be computed locally in the row and column decoders using $\boldsymbol{M}^{\mathsf{c}, (\ell-1)}_{i,:}$ (or $\boldsymbol{M}^{\mathsf{r}, (\ell-1)}_{i,:}$), the channel LLRs, and the iBDD-CR LUT.  

On the other hand, in BEE-SCC $|\bar{\bm{L}}^{(i),(\ell)}_{j,:}|$ is used to find the two least reliable bits of ${\bm{\Psi}}^{(i),(\ell)}_{j,:}$. Note that according to the derived DE in Section~\ref{DE_analysis}, $|\bar{\bm{L}}^{(i),(\ell)}_{j,:}|$ is the reliability of the iBDD-CR output $\bar{\bm{\Psi}}^{(i),(\ell)}_{j,:}$ (see Fig.~\ref{BEE-SCC}). However, in BEE-SCC we pragmatically employ it as the reliability measure for ${\bm{\Psi}}^{(i),(\ell)}_{j,:}$. The motivation is that BDD reveals also some information about its input, e.g., in the case of decoding failure it shows that the input is not within  distance $t$ of any codeword. Using $|\bar{\bm{L}}^{(i),(\ell)}_{j,:}|$ as the reliability measure for the error-and-erasure attempt has also a practical implication. $\bar{\bm{L}}^{(i),(\ell)}_{j,:}$ is computed inside the decoder by the first decoding attempt (see first branch of Fig.~\ref{BEE-SCC}), hence, there is no need to exchange any soft value between component decoders in order to erase the two least reliable bits for the second decoding attempt. We remark that, in principle, one can employ the BEE-PC architecture for SCCs as well. However, we found that such scheme yields minor performance improvement compared to iBDD-CR.

\subsection{Complexity}\label{Comments:Complexity}

A detailed complexity comparison between BEE-PC, BEE-SCC, the hybrid decoding algorithms  in \cite{She18b,sheikhTCOM19,Yia2019,She19,GabrieleSABMSR2019,optimal_decsheikh}, and  AD  \cite{Hag18} would require  a hardware implementation of each decoder, and comparing the resulting overall throughput and energy consumption. This implementation is beyond the scope of this paper, and thus, here we provide an algorithmic-level comparison. In the following, we first evaluate the contribution of the message exchange between component decoders of BEE-PC and BEE-SCC to the overall internal decoder data flow, as an essential metric for high-throughput systems \cite{staircase_frank}. We  then compare this and other implementation requirements of BEE-PC and BEE-SCC with those of the algorithms in \cite{Hag18,She18b,sheikhTCOM19,Yia2019,She19,GabrieleSABMSR2019,optimal_decsheikh} from an algorithmic-level perspective.

As explained in Section~\ref{RABMP_PC}, for the $i$th row decoding of a PC, BEE-PC outputs $\boldsymbol{M}^{\mathsf{r}, (\ell)}_{i,:}$ with elements in $\{0,\pm1\}$, and sends it to the column decoders. Therefore, it may seem at a first glance that the contribution of the message exchange of BEE-PC to the decoder data flow is higher than that of iBDD-CR and iBDD\textemdash both exchange only binary messages. However, in what follows, we  show BEE-PC can be implemented more efficiently than a plain ternary message passing between component decoders. Recall that the elements of $\boldsymbol{M}^{\mathsf{r}, (\ell)}_{i,:}$ are ternary because BDD or EDD may fail. In such case, all elements of $\boldsymbol{M}^{\mathsf{r}, (\ell)}_{i,:}$ are zero. If the row decoders can somehow indicate the failure to the column decoders (and vice versa), then the components of $\boldsymbol{M}^{\mathsf{r}, (\ell)}_{i,:}$ become binary. For the efficient implementation of BEE-PC, we propose to extend $\boldsymbol{M}^{\mathsf{r}, (\ell)}_{i,:}$ by one bit, i.e., $\boldsymbol{M}^{\mathsf{r}, (\ell)}_{i,:}$ is extended from a length-$n$ vector to a vector of length $n+1$, in which  the extra bit is $1$ if $\boldsymbol{M}^{\mathsf{r}, (\ell)}_{i,:}$ corresponds to BDD/EDD decoding. In this case, the other $n$ bits of $\boldsymbol{M}^{\mathsf{r}, (\ell)}_{i,:}$ take value in $\{0,1\}$ using the mapping  $+1\mapsto 0$ and $-1\mapsto 1$. The extra bit is $0$ if $\boldsymbol{M}^{\mathsf{r}, (\ell)}_{i,:}$ corresponds to BDD/EDD failure. In this case, the other $n$ bits of $\boldsymbol{M}^{\mathsf{r}, (\ell)}_{i,:}$ are $0$. With this simple bit extension and mapping, $\boldsymbol{M}^{\mathsf{r}, (\ell)}_{i,:}$ becomes binary, at the cost of sending only a single extra bit per component code. As the component code of a PC is typically long to reduce the error floor, the contribution of exchanging an extra bit to the decoder data flow is negligible. For instance, BEE-PC decoding of a PC with BCH component code of length  $255$ and $511$, entails an increase in message exchange of only $0.4\%$ and $0.2\%$, respectively, compared to iBDD-CR and conventional iBDD. 

As shown in Section~\ref{BEE-SCC_sec},  BEE-SCC only exchanges (binary) hard decisions between component decoders (see \eqref{fD_SCC} and Fig.~\ref{BEE-SCC}), hence, the contribution of message exchange of BEE-SCC to the decoder data flow is the same as that of iBDD-CR and conventional iBDD. 

Two other implementation aspects we consider here are the complexity entailed by EDD and the LLR calculations. BDD is usually implemented based on the Berlekamp-Massey algorithm \cite{Justesen2010}. EED can also be implemented by modifying the Berlekamp-Massey algorithm, hence, the complexity of EED  and BDD are roughly similar \cite{blahut_2003}. 
Both BEE-PC and BEE-SCC require to compute and store the channel LLRs for the code bits, and an  algorithm to find the two least reliable bits per component code. It is important to remark  that the memory required to store the channel LLRs static, as they are not updated during the decoding process. It is known that static memory is significantly less costly than dynamic memory in hardware implementation \cite{chris2019}.

In Table~\ref{Tabcomp}, we compare different features of AD \cite{Hag18}, iBDD-SR \cite{sheikhTCOM19}, SABM \cite{Yia2019}, iBDD-CR \cite{optimal_decsheikh}, SABM-SR \cite{GabrieleSABMSR2019}, iGMDD-SR \cite{She18b}, and BMP-GMDD \cite{She19}  with those of BEE-PC and BEE-SCC.\footnote{Table~\ref{Tabcomp} briefly compares the implementation requirements of different decoders. We refer the reader to \cite{Hag18,She18b,sheikhTCOM19,Yia2019,She19,GabrieleSABMSR2019,optimal_decsheikh}, for more extensive description of each decoding algorithm.} As can be seen from the table, some of the proposed algorithms are evaluated for PCs or for transmission over the bi-AWGN channel, however, the others are investigated for both PCs and SCCs and CM scheme. Furthermore, one can see that each algorithm requires different memory types and message passing. We highlight that the contribution of message passing to the decoder data flow is estimated based on BCH component code parameters $n=255$, $n=511$, and $t=\{2,3,4\}$, which are typically considered in the literature \cite{Hag18,She18b,sheikhTCOM19,Yia2019,She19,GabrieleSABMSR2019,optimal_decsheikh}.

\section{Conclusions}\label{Sec:Conclusions}

We extended the recently introduced iBDD-CR algorithm for PCs to SCCs via a DE analysis of the SC-GLDPC code ensemble encompassing SCCs as particular instances. We further proposed  two novel decoding algorithms for PCs and SCCs which augment iBDD-CR by introducing a second decoding attempt based on EED of the component codes. Both BEE-PC and BEE-SCC offer sizeable performance gains   compared to standard iBDD, up to  $0.88$ dB for a BICM system using $256$-QAM. These translate into an enhancement of the optical reach of up to 33\%. Moreover, the internal decoder data flow of BEE-PC and BEE-SCC resulting from the exchange of messages between component decoders is roughly the same of that of (standard) iBDD, which makes the proposed algorithms excellent candidates for next generation ultra high-throughput systems.


Future work includes hardware implementation of BEE-PC and BEE-SCC  and measuring the corresponding energy consumption per information bit at high throughputs. 

%

\section*{Acknowledgment}

The authors would like to thank Dr. Christian H\"ager for  providing the simulation results of AD in Figs.~\ref{simPCbi}, \ref{sim1}, and \ref{sim2}. 

\balance



\end{document}